\documentclass[prl,twocolumn,superscriptaddress,floatfix,preprintnumbers,amssymb,amsmath]{revtex4-2}

\usepackage{hyperref}
\usepackage{graphicx}
\usepackage{xcolor}
\usepackage{svg}
\usepackage{printlen}
\usepackage{layouts}
\usepackage{siunitx}

\begin{document}

\title{Demonstration of dual Shapiro steps in small Josephson junction}

\author{Fabian~Kaap}
\email{fabian.kaap@ptb.de}
\affiliation{Physikalisch-Technische Bundesanstalt, Bundesallee 100, 38116 Braunschweig,
	Germany}
\affiliation{RWTH Aachen University, 52056 Aachen, Germany}

\author{Christoph~Kissling}

\author{Victor~Gaydamachenko}

\author{Lukas~Grünhaupt}

\affiliation{Physikalisch-Technische Bundesanstalt, Bundesallee 100, 38116 Braunschweig, Germany}

\author{Sergey~Lotkhov}

\affiliation{Physikalisch-Technische Bundesanstalt, Bundesallee 100, 38116 Braunschweig,
	Germany}

\begin{abstract}
	
	Bloch oscillations in small Josephson junctions were predicted theoretically as the quantum dual to Josephson oscillations. A significant consequence of this prediction is the emergence of quantized current steps, so-called dual Shapiro steps, when synchronizing Bloch oscillations to an external microwave signal. These steps potentially enable a fundamental standard of current $I$, defined via the frequency $f$ of the external signal and the elementary charge $e$, $I=\pm n \times 2ef$, where $n$ is a natural number. Here, we realize this fundamental relation by synchronizing the Bloch oscillations in small Al/AlO$_\mathrm{x}$/Al  Josephson junctions to sinusoidal drives with frequencies varying from 1 to 6\:GHz and observe dual Shapiro steps up to $I\approx \SI{3}{\nano \ampere}$. 
	Inspired by today’s voltage standards and to further confirm the duality relation, we investigate a pulsed drive regime, which is dual to the single flux quantum mode of Josephson oscillations, and observe a similar asymmetric pattern of dual Shapiro steps. 
	This work confirms quantum duality effects in Josephson junctions and paves the way towards a range of applications in quantum metrology based on well-established fabrication techniques and straightforward circuit design.

\end{abstract}
\maketitle
	
Today’s voltage standard provided by national metrology institutes \cite{Jeanneret09} relies on the synchronization of Josephson oscillations \cite{Josephson62} with an external rf-drive. This leads to equidistant voltage steps, the so-called Shapiro steps \cite{Shapiro63}. The dual effect of quantized current steps was predicted theoretically almost 40 years ago \cite{Likharev85,Averin85} and originates from the quantum mechanical description of small Josephson junctions, where charge and phase are conjugate variables. Quantized currents based on this duality could potentially close the metrological triangle, uniting the ampere, the volt and the ohm through fundamental constants and frequency. The closure of the triangle would open the way for combining macroscopic quantum relations of voltage, current, and resistance in one device and allow to investigate the consistency of the three relations. In terms of implementing a quantum current standard, significant advantages of using Josephson junctions for this purpose, as compared to e.g. semiconductor pumps \cite{Stein16,Giblin12}, are higher frequencies in the GHz-range and technological compatibility with Josephson voltage standards.

To observe dual Shapiro steps, the Josephson junction has to be embedded in a high impedance environment $Z_\mathrm{env} > R_\mathrm{Q} = h/4e^2 \approx \SI{6.4}{\kilo\ohm}$ in order to suppress quantum fluctuations of charge and enable the observation of charging effects. First indications of dual Shapiro steps were reported a few years following the original prediction \cite{Kuzmin91} and relied on a purely resistive environment. However, strong heating effects made it difficult to observe quantized current steps \cite{Kuzmin94}, such that no detailed study of the synchronization mechanism could be carried out. To solve this issue a biasing circuit combining high-ohmic off-chip resistors and compact on-chip inductances \cite{Arndt18}, whose specific impedance exceeds the quantum resistance was proposed. Such so-called superinductances can in principle be realized via Josephson junction chains \cite{Manucharyan09,Pop14,Pechenezhskiy20,Crescini23}, nanowires on membranes \cite{Peruzzo20,Peruzzo21} or high kinetic inductance leads \cite{Gruenhaupt18,Maleeva18,Gruenhaupt19,Kamenov20,Moshe20,Shaikhaidarov22,Torras24,Kaap22} of typically a few tens of µm length. Such elements together with small Josephson junctions allow to use capacitive coupling for synchronization of Bloch oscillations (BO) and thus, mitigate rf-heating in the resistors. 

Here, we utilize the large resistivity of oxidizied titanium \cite{Lotkhov13} and the kinetic inductance of granular aluminium \cite{Rotzinger16,Gruenhaupt19} to implement the high impedance environment and combine it with standard Al/AlO$_\mathrm{x}$/Al Josephson junction fabrication technique \cite{Dolan77} to realize an experiment guided by the theoretical proposals \cite{Likharev85,Averin85,Arndt18}. We demonstrate pronounced quantized current steps in small Josephson junctions by synchronizing their BO to external microwave signals between 1 and 6\:GHz, without relying on discrete high-quality resonant modes, which limited recent previous evidence of dual Shapiro steps to four distinct frequencies \cite{Crescini23}. Using shadow evaporated Josephson junctions, we provide an alternative to coherent quantum phase slip circuits, based on advanced fabrication techniques \cite{Shaikhaidarov22}, which demonstrated the first pronounced quantized current steps overcoming the heating challenge of initial experiments \cite{Kuzmin91,Kuzmin94}.
To further investigate the duality relation, we implement pulsed external signals for synchronization of BO, a technique to enhance the width of quantized voltage steps \cite{Maibaum11,Benz96}. We demonstrate an asymmetric pattern of dual Shapiro steps consistent with theoretical predictions \cite{Maggi96}, reinforcing the observed effect as the dual to quantized voltage steps.

\begin{figure*}[]
	\centering\includegraphics[width=1\textwidth]{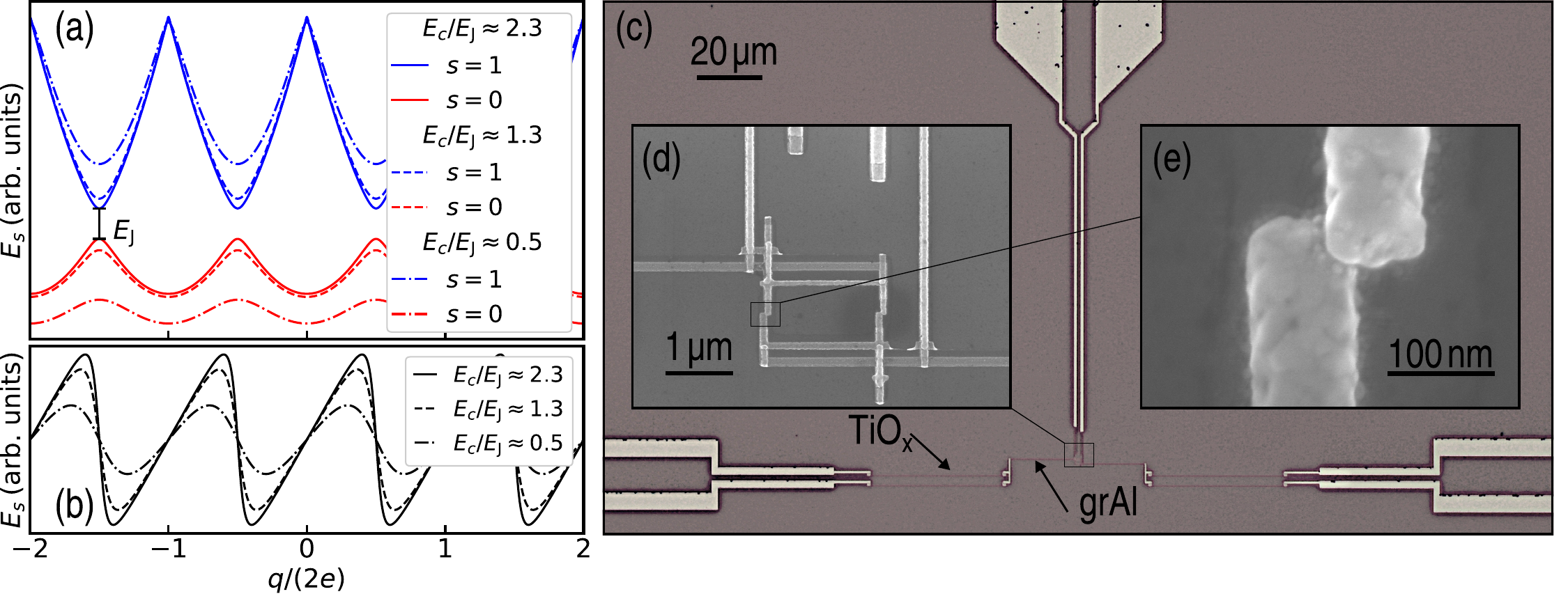}
	\caption{Bloch oscillations: energy diagram and experimental realization. (a) Energy $E_s(q)$ of the two lowest Bloch bands for different ratios $E_C/E_\mathrm{J}$. (b) Ground state oscillations of voltage calculated by taking the derivative $\frac{\partial E_0}{\partial q}=V$. These so-called Bloch oscillations have a frequency $f_\mathrm{B} \approx I_\mathrm{bias}/2e$ (see main text). (c-e) Optical and scanning electron microscope images of a representative device consisting of a dc-SQUID [see panel(d)] with two Josephson junctions of area $A_\mathrm{JJ} \approx \qtyproduct[product-units = bracket-power]{20 x 30}{\nano\metre}$ [see panel(e)] embedded in a bias circuitry. The slot line from the top delivers the rf-driving signal via a capacitive coupling with $C_\mathrm{c} \approx \SI{0.3}{\femto\farad}$ while the horizontal leads are used for four-probe dc-measurements. The bias circuitry includes the \SI{20}{\micro\metre}-long granular aluminium inductor \cite{Gruenhaupt19} with kinetic inductance $L_\mathrm{grAl} \approx \SI{50}{\nano\henry}$ and characteristic impedance $Z_\mathrm{grAl} \approx 10.5\mathrm{k\Omega} > R_\mathrm{Q}= h/4e^2  \approx \SI{6.4}{\kilo\Omega}$ and the \SI{40}{\micro\metre}-long  $\mathrm{TiO_x}$ resistors \cite{Lotkhov13,Kaap24} of $R_\mathrm{TiO_x}\approx \SI{275}{\kilo\ohm}$. This provides a high impedance environment $Z_\mathrm{env} \ge R_\mathrm{Q} $, suppresses quantum fluctuations of charge and enables Bloch oscillations \cite{Likharev85}.}
	\label{Fig0}
\end{figure*}

\begin{figure}[]
	\centering
	\includegraphics[width=1\linewidth]{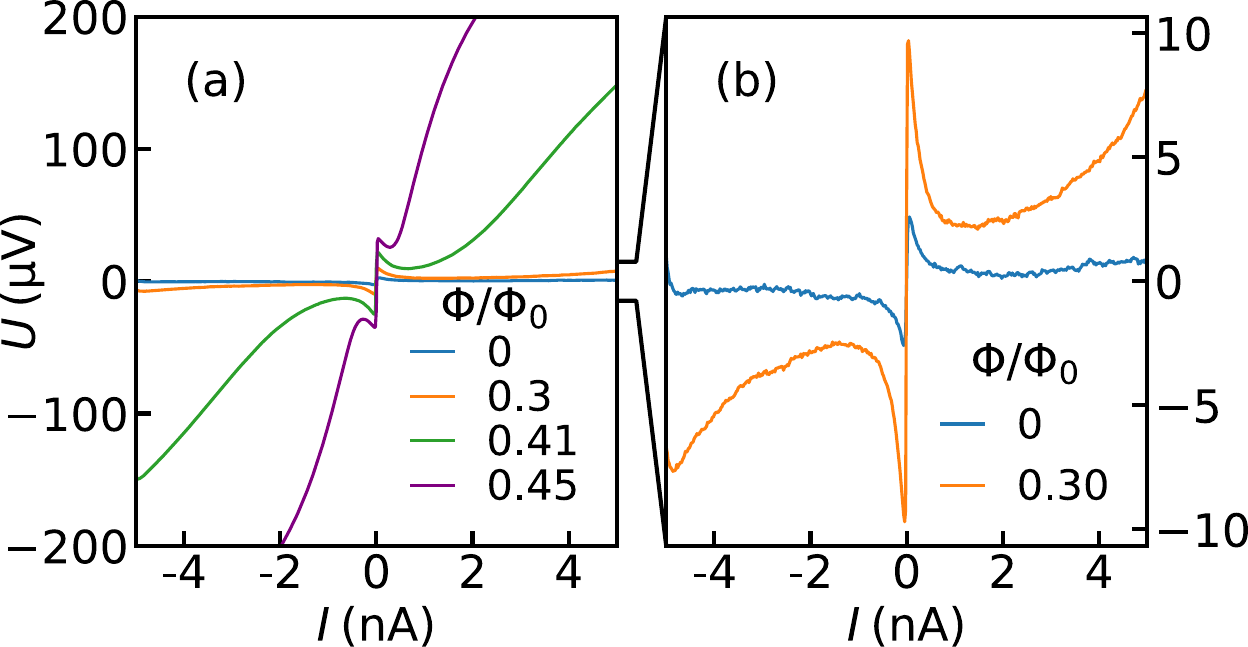}
	\caption{External flux dependent Coulomb blockade of the device. (a) $IV$-curves of the sample [cf. Fig.~\ref{Fig0}(c-e)] for external flux values $0\le \Phi/\Phi_0 \le 0.45$. By approaching $\Phi_0/2$ and thus increasing the ratio $E_C/E_\mathrm{J}$ we observe wider ranges of the Coulomb blockade and steeper slopes of the resistive branch caused by Landau-Zener tunnelling (LZT) \cite{Zaikin88,Geigenmuller88,Kuzmin96} (see main text). (b) Zoom-in on the $IV$-curves for $\Phi/\Phi_0 = 0,0.3  $. A back-bending to almost zero voltage can be observed, which is a hallmark of Bloch oscillations in a dc-measurement. We identify flux values $0 \le \Phi/\Phi_0 \le 0.3 \: (1.3 \le E_C/E_\mathrm{J} \le 2.3)$ as a suitable range for observing dual Shapiro steps (see Supplementary Fig.~7).}
	\label{Fig1}
\end{figure}

The behaviour of a Josephson junction is determined by its charging energy $E_C=\frac{e^2}{2C}$ and its Josephson energy $E_\mathrm{J}=\frac{\Phi_0 I_\mathrm{c}}{2\pi}$, where $e,\Phi_0$ are the electron charge and the flux quantum and $C$, $I_\mathrm{c}$ are the capacitance and critical current of the Josephson junction. Taking into account that the charge and phase operators are governed by the canonical commutation relation $[ \hat{Q}, \hat{\varphi}]=2ei$, the Hamiltonian is given by

\begin{equation}
	\hat{H} = \frac{\hat{Q}^2}{2C} - E_\mathrm{J} \cos (\hat{\varphi}).
	\label{eq_Hamiltonian}
\end{equation}

This Hamiltonian can be solved using Blochs theorem \cite{Bloch29,Kittel05} and yields periodic energy bands \mbox{$E_{s}(q)=E_{s}(q+2e)$}, where $s \in \mathbb{N}_0$ is the level number and $q$ is the so-called quasicharge, which represents the charge injected to the junction \cite{Averin85,Likharev85}.

In Fig.~\ref{Fig0}(a) the two lowest energy bands ($s=0,1$) are shown for different ratios $E_C/E_\mathrm{J}$. In the ground-state approximation, the dynamics of this system is given by the so-called Langevin equation:

\begin{equation}
	\dot q = I_\mathrm{bias}-V[q(t)]/R + I_{\rm {rf}}(t) + \tilde{I}(t),
	\label{Langevin_rf}
\end{equation}
with $I_\mathrm{bias}$ a dc bias current, $V=\frac{\partial E_{0}}{\partial q}$, $R$  the real part of the impedance of the environment, which we for simplicity assume to be purely ohmic, $\tilde{I}(t)$ the fluctuation current and $I_\mathrm{rf}$ an external rf-signal \cite{Likharev85,Averin85}. If $I_{\rm {rf}}(t)=0$, $\tilde{I}(t)=0$ and $I_\mathrm{bias}<\max[\frac{\partial E_{0}}{\partial q}]/R$, there is a stationary solution for Eq.~\eqref{Langevin_rf}, a Coulomb blockade. For currents $I_\mathrm{bias}>\max [\frac{\partial E_{0}}{\partial q}]/R$ the solution is periodic in time and results in BO with frequency $f \approx I_\mathrm{bias}/2e$ [see Fig.~\ref{Fig0}(b)]. These BO can synchronize to an external rf-current $I_\mathrm{rf}$, generating quantized current steps at $I=\pm n\times 2ef$, referred to as dual Shapiro steps, where $n$ is a natural number.

\begin{figure*}[t]
	\centering\includegraphics[width=\textwidth]{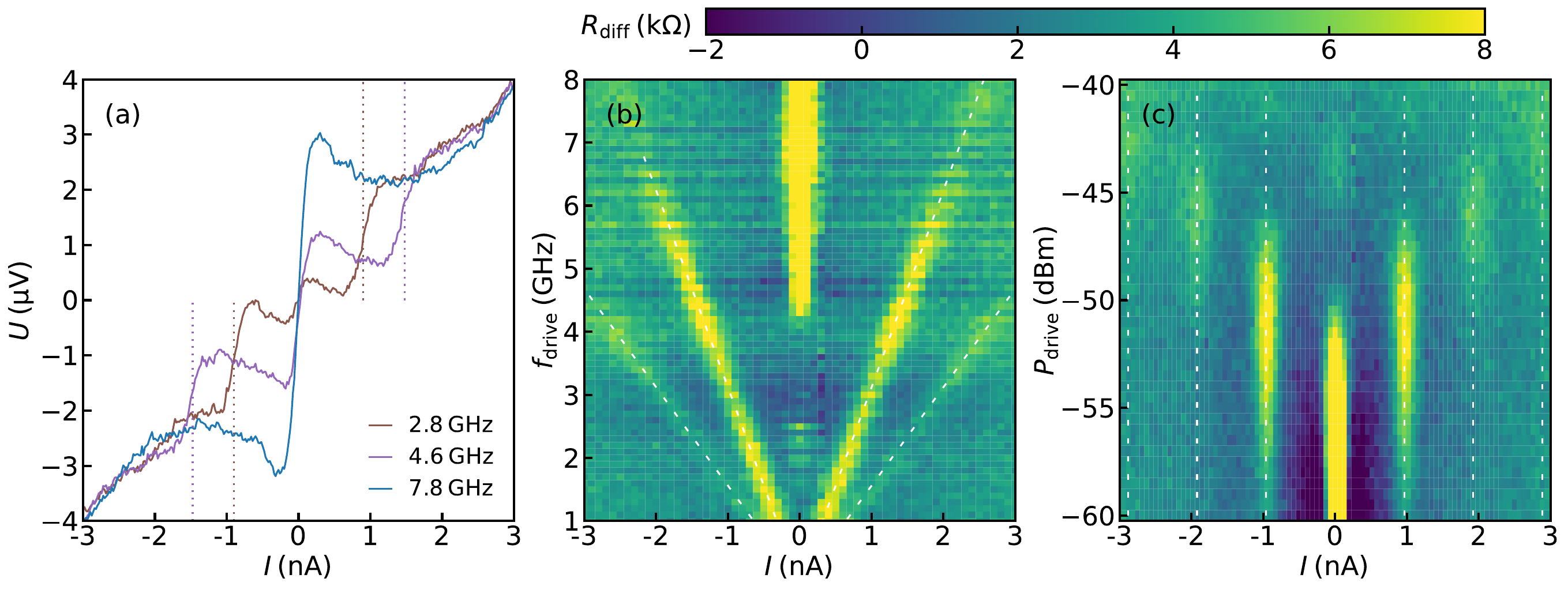}
	\caption{Dual Shapiro steps due to sinusoidal rf-drive. (a) Measured $IV$-curve with $f_\mathrm{drive}=2.8\mskip\thinmuskip \unit{\GHz} \: (\mathrm{brown}), \:4.6\mskip\thinmuskip \unit{\GHz} \:$ \\$ (\mathrm{violet}), \:7.8\mskip\thinmuskip \unit{\GHz} \: (\mathrm{blue}) $  for rf-power at the chip input of $p_\mathrm{drive}=\SI{-49}{\deci\bel m}$. The dashed lines indicate the positions of the first dual Shapiro steps according to $I_\mathrm{dSs}=\pm 2ef_\mathrm{drive}\approx \pm 0.93 \: \mskip\thinmuskip \unit{\nano\ampere}$ and $\pm\SI{1.53}{\nano\ampere}$. (b) Differential resistance $R_\mathrm{diff}= \frac{dU}{dI}$ measured using a lock-in amplifier for different drive frequencies $f_\mathrm{drive}$ at constant drive power $p_\mathrm{drive}=\SI{-49}{\decibel m}$. Dashed white lines confirm the expected position of dual Shapiro steps according to $I_\mathrm{dSs}=\pm n \: 2ef_\mathrm{drive}$. (c) $R_\mathrm{diff}$ for different powers $p_\mathrm{drive}$ at fixed frequency $f_\mathrm{drive}=\SI{3}{GHz}$, leading to steps at $I_\mathrm{dSs}\approx \pm n\cdot \SI{0.96}{\nano\ampere}$. The amplitude of the Coulomb blockade and the dual Shapiro steps vary with $p_\mathrm{drive}$ (see main text).}
	\label{Fig2}
\end{figure*}

Fig.~\ref{Fig0}(c-e) shows optical and scanning electron microscope (SEM) pictures of a representative device. We fabricate the chip using three lithographic steps on a thermally oxidized silicon wafer \cite{Kaap22}. First, we pattern a wiring structure from Au (not shown) using optical lift-off lithography. Subsequently, we fabricate \SI{150}{\nano\metre} wide and \SI{15}{\nano\metre} thick TiO$_\mathrm{x}$ resistors \cite{Lotkhov13}. Finally, we in-situ combine a \SI{20}{\nano\metre} thick granular aluminium superinductor \cite{Gruenhaupt19} with double-angle evaporation Al/AlO$_\mathrm{x}$/Al Josephson junctions \cite{Dolan77}. To implement the last lithographic step we utilize a PMMA/copolymer liftoff mask with a controllable undercut \cite{Lecocq11}. Each of the two Josephson junctions of the SQUID has an area $A_\mathrm{JJ} \approx \qtyproduct[product-units = bracket-power]{20 x 30}{\nano\metre}$ [see Fig.~\ref{Fig0}(e)] leading to an upper bound of the charging energy $E_C=e^2/(4 A_\mathrm{JJ} \times \SI{50}{\femto\farad\micro \meter^{-2}}) = \SI{1.3}{\milli\eV}$. The true experimental value of $E_C$, however, includes the full 3D junction topology as well as the stray capacitances of the leads, and is measured to be $E_C \approx \SI{80}{\micro\eV}$ (see Supplementary Fig.~3). In order to keep the Josephson energy $E_\mathrm{J}$ comparable to the charging energy, we dynamically oxidize at $P_\mathrm{O_2} = \SI{0.1}{\pascal}$ for $\SI{300}{s}$, resulting in $E_\mathrm{J} \approx \SI{60}{\micro\eV}$ (see Supplementary Fig.~3). The leads of the SQUID are connected to granular aluminium strips, which split up in two oxidized titanium resistors \cite{Lotkhov13,Kaap24} [see Fig.~\ref{Fig0}(c)]. From a measurement of the leads with the cryostat at base temperature, we extract the resistance of the TiO$_\mathrm{x}$ strips $R_\mathrm{TiO_X}=\SI{275}{\kilo\ohm}$. Using the Matthis-Bardeen formula we estimate $L_\mathrm{grAl}= 0.18\hbar R /(k_\mathrm{B} T_\mathrm{c})\approx \SI{50}{\nano\henry}$ \cite{Rotzinger16}, from a room temperature resistance measurement of the grAl strip. Aside from their inductive contribution leading to an characteristic impedance $Z_\mathrm{grAl}>R_\mathrm{Q}$, the grAl strips might act as quasiparticle filters due to their larger superconducting energy gap as compared to the Al films \cite{Sun12}. Quasiparticle tunneling is a process that partially replaces the coherent Cooper pair tunnelling associated with Bloch oscillations. Reducing the presence of quasiparticles in the vicinity of the Josephson junction will thus result in a larger Coulomb blockade. An external rf-signal for synchronization of the BO is supplied via a \SI{50}{\ohm} slot line and capacitively couples to the SQUID [see Fig.~\ref{Fig0} (c-d)].

\section{Results}

All measurements were performed in a dilution refrigerator at $\sim \SI{15}{\milli\kelvin}$. All dc lines include three meter long Thermocoax$\textsuperscript{TM}$ cable filters \cite{Zorin95} connecting the room temperature electronics to the mK sample stage. The rf driving signals applied to the sample are attenuated by \SI{50}{\decibel} at different temperature stages of the dilution cryostat (see Supplementary Fig.~1). The rf-powers quoted in the following refer to the power level at the mK-stage, where additional cable losses are neglected and should be equal for all experiments.

\subsection{Coulomb blockade}
Fig.~2(a) shows the $IV$-curves of the sample without external rf-drive for values of the external flux between $0\le \Phi/\Phi_0 \le 0.45$. As can be seen, the effective Josephson energy of the dc-SQUID $E_\mathrm{J}(\Phi)=E_\mathrm{J,0} | \cos (\pi \frac{\Phi}{\Phi_0})|$, therefore the shape of the Bloch bands and, consequently, the $IV$-curve are tuned. Close to  $\Phi \approx {\Phi_0}/{2}$ the Coulomb blockade increases, which can be explained by the increasing amplitude of the first Bloch band [cf. Fig.~\ref{Fig0}(a)]. Moreover, for large bias currents $I_\mathrm{bias}$ the probability to tunnel from $E_{0}$ to $E_{1}$ increases and the so-called Landau-Zener-tunneling (LZT) \cite{Zaikin88,Geigenmuller88,Kuzmin96} destroys the coherence of the BO. The LZT manifests in a resistive branch in the $IV$-characteristics as a positive slope at currents exceeding an LZT onset current of a few nA. Since the gap between the lowest two bands is approximately proportional to $E_\mathrm{J}$, the onset current for this LZT branch rapidly decreases with the external flux and reaches a minimum at $\Phi/\Phi_0 = 0.5$. From measurements at different cryostat temperatures, we estimate an effective electron temperature of \SI{40}{\milli\kelvin} (see Supplementary Fig.~6). Given this temperature $k_\mathrm{B}T \ll  \Delta^{(0)} = \mathrm{min}(E_{1}) - \mathrm{max} (E_{0})$ the dynamics of the system are confined to the lowest energy band.

Fig.~\ref{Fig1}(b) shows a zoom-in of the $IV$-curves for $\Phi/\Phi_0 = 0$ and $0.3$. In the blue curve a clear backbending from the initial Coulomb blockade of $\sim\SI{4}{\micro\volt}$ down to $U\approx 0$ can be observed, indicating  negligible LZT up to currents of $\sim \pm \SI{5}{\nano \ampere}$. In comparison, the orange curve shows an extended blockade range of  $\sim\SI{18}{\micro\volt}$, but at the same time a reduced onset current of LZT of $\sim \SI{2}{\nano\ampere}$. As a compromise between maximizing the Coulomb blockade width, which sets an upper limit for the dual Shapiro step width, and the range of quantized currents limited by LZT, all measurements shown in this work are done at $\Phi/\Phi_0 = 0.3$, resulting in $E_C/E_\mathrm{J} \approx 2.3$ (see Supplementary Fig.~7).

\begin{figure*}[t]
	\centering\includegraphics[width=\textwidth]{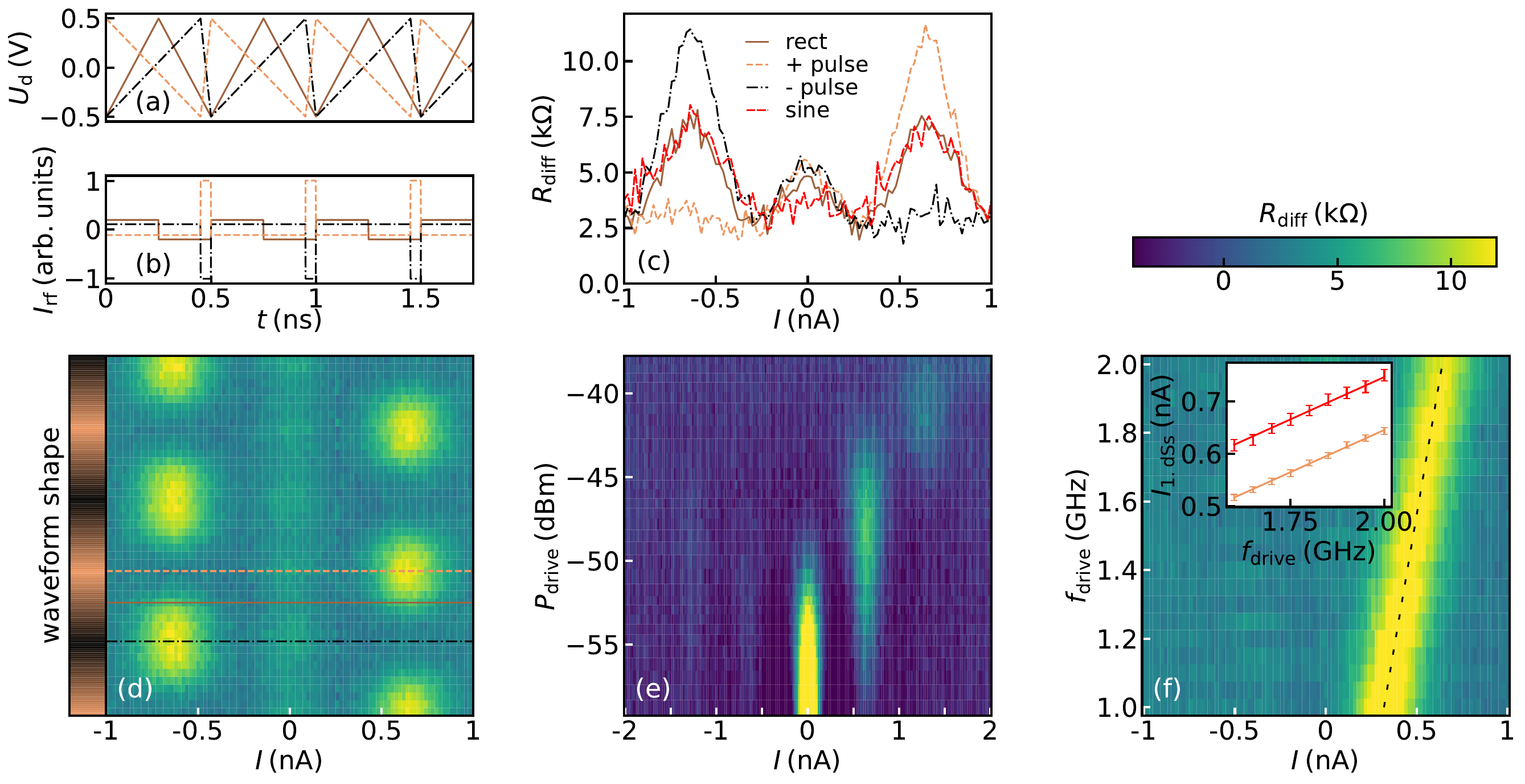}
	\caption{Shaping dual Shapiro steps by using different waveforms. (a) Dedicated waveforms $U_\mathrm{d}(t)$ used to drive the dual Shapiro steps. (b) rf-current injection according to $I_\mathrm{rf} \propto C_\mathrm{c} \times\frac{dU_\mathrm{d}}{dt}$. The symmetric waveform (brown) leads to a rectangular current profile, while the negative (positive) sloped sawtooth signal lead to short positive (negative) current pulses (black and beige, respectively). (c) Differential resistance $R_\mathrm{diff}$ for the three current profiles shown in (b) and a sinusoidal drive (red) with $f_\mathrm{drive}=\SI{2}{GHz}$ and $p_\mathrm{drive}=\SI{-48}{\decibel m}$ (see Supplementary Fig.~4). Depending on the pulse, either the positive or negative dual Shapiro step is enhanced and reaches almost twice as high peak values $R_\mathrm{diff}$ as compared to the symmetric drives (triangular and sinusoidal), while the opposite step is suppressed. (d) $R_\mathrm{diff}$ measured for linearly sweeping back and forth between a positive (beige) and negative pulsed drive (black). The colorbar on the y-axis indicates the gradual change of the sawtooth polarity. Colored lines mark the curves shown in (c). (e) Power dependence of $R_\mathrm{diff}$ using a pulsed drive at $\SI{2}{\giga\hertz}$. Increasing the power of the pulse leads to a gradual activation of the consecutive dual Shapiro steps. (f) Differential resistance $R_\mathrm{diff}$ for different frequencies $f_\mathrm{drive}$ of the pulsed drive. Over the whole frequency range the position of the positive dual Shapiro step (black dashed line) aligns with $I_\mathrm{dSs}=2ef_\mathrm{drive}$, while the negative step is suppressed. Inset: Measured current of the first dual Shapiro step as a function of the drive frequency for the sinusoidal drive (red and plotted with \SI{0.1}{\nano\ampere} offset) and the pulsed drive (beige). From a linear fit we can extract the charge of an electron according to $I=2ef_\mathrm{drive}$. For the sinusoidal drive we extract $e_\mathrm{sine}=(1.612\pm0.010)\times 10^{-19}\:\unit{\coulomb}$ and for the pulsed drive $e_\mathrm{pulse}=(1.610\pm0.005)\times 10^{-19}\:\unit{\coulomb}$. The errorbars correspond to the standard deviation of the step position extracted by a least squares fit of a Gaussian (see Supplementary Fig.~4). Note that the errorbars for the pulsed drive are smaller.}
	\label{Fig5}
\end{figure*}

\subsection{Dual Shapiro steps using sinusoidal drive}

In the following we perform measurements with a sinusoidal drive applied to the SQUID at different GHz-frequencies and rf-power levels. Fig.~\ref{Fig2}(a) shows the $IV$-curves measured at a drive power of $p_\mathrm{drive}= \SI{-49}{\decibel m}$ and frequency of $f_\mathrm{drive}=2.8\mskip\thinmuskip \unit{GHz}$ and $\SI{4.6}{\giga\hertz}$ featuring two distinct steps around the quantized values of $I=\pm 2ef_\mathrm{drive}\approx 0.93 \: \mskip\thinmuskip \mathrm{nA}$ and \SI{1.53}{\nano\ampere}, respectively (brown and violet dotted lines). The Coulomb blockade as well as the dual Shapiro steps are skewed as compared to the non-rf case [see orange curve in Fig.~\ref{Fig1}(b)], which we attribute to additional heating caused by the external rf-drive. We estimate an electron temperature of $\sim \SI{250}{\milli\kelvin}$, which is a factor of $\sim6$ higher than for the undriven measurements of the Coulomb blockade (see Supplementary Fig.~6). The continuous raise of voltage outside the range $|I| > \SI{2}{\nano\ampere}$ is caused by the LZT and resembles the LZT branch in the orange curve in Fig.~\ref{Fig1}(b).

Fig.~\ref{Fig2}(b) shows the differential resistance $R_\mathrm{diff}= \frac{dU}{dI}$ measured with a lock-in amplifier for different drive frequencies $f_\mathrm{drive}$ at a constant power $p_\mathrm{drive}= \SI{-49}{\decibel m}$. The white dashed lines indicate the due positions of the first and second dual Shapiro step, $I_\mathrm{peak}= \pm n\times 2ef_\mathrm{drive}$. Up to frequencies of $\sim \SI{6}{\giga\hertz}$ the first step can clearly be identified, while towards higher frequencies the steps are vanishing [see Fig.~\ref{Fig2}(a)] and the Coulomb blockade reappears, which indicates lower rf-power reaching the SQUID. We attribute this to stronger reflection of the rf-signal at higher frequencies in the slot line. Additionally, for synchronization at higher frequencies larger power is needed, since the width of (dual) Shapiro steps as a function of power scales proportional to $|J_n(\alpha)|^b$, where $J_n$ are the Bessel functions of the first kind, $\alpha \propto \sqrt{P_\mathrm{drive}}/f_\mathrm{drive}$ is a normalized power and $b$ is an exponent, which depends on the impedance of the environment \cite{Kurilovich24,Shaikhaidarov22}.

Fig.~\ref{Fig2}(c) shows the dependence of $R_\mathrm{diff}$ for rf-power between $\SI{-60}{\decibel m} \le p_\mathrm{drive} \le \SI{-40}{\decibel m}$ at a fixed drive frequency $f_\mathrm{drive}=\SI{3}{\giga \hertz}$. The equidistant set of peaks in $R_\mathrm{diff}(I)$ indicates the dual Shapiro steps separated by $\Delta I = \SI{0.96}{nA}$. At low power, $p_\mathrm{drive}\approx \SI{-60}{\decibel m}$, the Coulomb blockade peak dominates, while the first dual Shapiro steps are barely visible. By increasing $p_\mathrm{drive}$ the first current steps grow larger, as indicated by the higher resistance peaks, and they reach the maximum peak height at $p_\mathrm{drive}=\SI{-51}{\decibel m}$, while the Coulomb blockade disappears. Up to the higher power levels around $p_\mathrm{drive}\approx \SI{-47}{\decibel m}$ the second, and around \SI{-43}{\decibel m} the third dual Shapiro steps can be seen. Higher-order dual Shapiro steps are not visible due to LZT destroying the coherence of the Bloch oscillations. This measured pattern of dual Shapiro steps can be modelled using Eq.~(1) and qualitatively reproduces the data obtained in Refs.~\cite{Shaikhaidarov22,Maggi96,Tien63} (see Supplementary Fig.~5). We note that for higher rf-power the increased heating leads to smearing of the features. As mentioned before, the power dependence follows the Bessel-like behaviour. However, our data does not allow to extract the exact exponent $b$ of this dependence.

\subsection{Pulse-driven dual Shapiro steps}

To implement a pulsed drive for the SQUID we apply a sawtooth signal $ U_{\rm d}(t) $ to the slot line, which yields $ I_{\rm {rf}} \propto C_{\rm c} \times dU/dt $. Fig.~\ref{Fig5}(a) depicts the signal profile of three different waveforms, namely a positive sawtooth, a negative sawtooth and a symmetric triangular waveform. The derived rf-current through the SQUID is sketched in Fig.~\ref{Fig5}(b). For the triangle we assume a symmetric rectangular current profile of low amplitude, while both sawtooth waveforms will generate a unipolar current pulse of larger amplitude. In Fig.~\ref{Fig5}(c) the differential resistance $R_\mathrm{diff}$ is shown for the three current profiles displayed in Fig.~\ref{Fig5}(b). For comparison, we plot similar data obtained for an optimized sinusoidal drive with $f_\mathrm{drive}=\SI{2}{\giga\hertz}$ and $p_\mathrm{drive}=\SI{-48}{\decibel m}$ (see Supplementary Fig.~4). For the triangular waveform both positive and negative branch exhibit a similar peak height as in the case with a sinusoidal drive. In contrast, if using a sawtooth-like waveform generating current pulses, the peak height of $R_\mathrm{diff}$ of one of the two opposite steps is approximately doubled in magnitude, while the other step is suppressed (see Supplementary Fig.~5, cf. Ref.~\cite{Maggi96}). As can be seen in Fig.~\ref{Fig5}(d), by changing the drive waveform between a positive and a negative sawtooth (pulses) [beige and black curves in Fig.~\ref{Fig5}(a,b)], we observe a modulation of measured curves between the three representative cases shown in Fig.~\ref{Fig5}(c), which are indicated by the three colored lines in panel (d). We note, that the largest differential resistance $R_\mathrm{diff,peak}$ of the measured steps occurs for maximally tilted sawtooth signals leading to the highest current pulse in Fig.~\ref{Fig5}(b).

The power dependence of $R_\mathrm{diff}$ measured with pulsed driving at \SI{2}{\giga\hertz} is shown in Fig.~\ref{Fig5}(e). By increasing the power, the first dual Shapiro step appears at \SI{-50}{\decibel m}, where the Coulomb blockade fades. Further increasing the power above \SI{-48}{\decibel m} decreases the first dual Shapiro step and the second step appears at \SI{-43}{\decibel m}. This power dependence of the individual steps is in qualitative agreement with numerical simulations (see Supplementary Fig.~5, cf. Ref.~\cite{Maggi96}). Note that to apply drive powers $P_\mathrm{drive}>\SI{-44}{dBm}$, the total attenuation in the lines is reduced to \SI{40}{dB}.

In Fig.~\ref{Fig5}(f) we plot $R_\mathrm{diff}$ for pulsed drives at different frequencies up to \SI{2}{\giga\hertz}, showing that the asymmetry in the $IV$-curve persists and the position of the step follows the fundamental relation $I=2ef_\mathrm{drive}$ (black dashed line). To quantify the enhancement of the quantized current step due to the pulsed drive [see inset of Fig.~\ref{Fig5}(f)], we extract the position of the first dual Shapiro step and compare it to the case of a sinusoidal drive (see Supplementary Fig.~4, shown with \SI{0.1}{\nano\ampere} for visualization). The power of the sinusoidal drive, $p_\mathrm{drive} = \SI{-48}{\decibel m }$, was set to maximize the step width and is comparable to the power of the sawtooth drives with \SI{-47.8}{\decibel m}. The extracted value of the electron charge for the pulsed drive $e_\mathrm{pulse}=(1.610\pm0.005)\times 10^{-19}\:\unit{\coulomb}$ is showing smaller statistical errors than for the sinusoidal case $e_\mathrm{sine}=(1.612\pm0.010)\times 10^{-19}\:\unit{\coulomb}$.

\section{Discussion}

In conclusion, we experimentally show the occurrence of Bloch oscillations in Josephson junctions embedded in a high impedance environment made from granular aluminium and oxidized titanium. By synchronizing the Bloch oscillations with an external sinusoidal rf-drive, we demonstrate pronounced dual Shapiro steps and quantized currents up to \SI{3}{\nano\ampere}. Using pulsed drives, we show an asymmetric pattern of the dual Shapiro steps. This research serves as a stepping stone for future metrological applications, namely the development of new quantum standards for currents in the nA-range and thus, for closing the metrological triangle. For future experiments, the detailed studies of dual Shapiro steps arising due to a wide variety of driving waveforms appear to be of particular interest. Combining the demonstrated technique of drives with further improvements in the circuit layout like an optimized rf-line geometry and using even smaller, point-like contacts could lead to even wider and more precise current steps. With those improvements, it might be possible to reach the current metrological precision of current standards realized via single electron pumps ($\approx \SI{0.2}{ppm}$ \cite{Stein16}).

During the review of our manuscript we became aware of a preprint by Antonov et al. \cite{Antonov24}, demonstrating dual Shapiro steps in Al/AlO$_\mathrm{x}$/Al Josephson junctions. However, we note that our result seem to contradict the observation of Antonov et. al that dual Shapiro steps are only observed in samples with a Coulomb blockade smaller than \SI{5}{\micro\volt}.

\section{Methods}
\subsection{Device fabrication}

The total device fabrication consists of three lithographic steps on a \SI{380}{\micro\meter} thick silicon substrate with a \SI{600}{\nano\meter} thermal oxide layer. All metal layers are structured by lift-off. We thermally evaporate a \SI{45}{\nano\meter} thick gold layer on top of a \SI{5}{\nano\meter} titanium layer for improved adhesion. This first wiring structure is patterned by photolithograpy using a Süss MA6 contact aligner and contains bond pads to connect the chip to the sample holder. The two following lithography steps for the TiO$_\mathrm{x}$ resistors and the Al structures utilize a commercial \SI{100}{\kilo\volt} electron beam writer. For these depositions we use a custom made electron beam evaporation system consisting of an evaporation chamber and a substrate chamber, in which the substrate can be tilted for shadow evaporation. To create the TiO$_\mathrm{x}$ resistors we evaporate titanium at a rate of \SI{2}{\angstrom\second^{-1}}, while the oxygen pressure inside the substrate chamber is kept at $\SI{3e-6}{\milli\bar}$. In order to enable a good electrical contact with the following layers the ends of the TiO$_\mathrm{x}$ resistors are sealed with a \SI{25}{\nano\meter} AuPd layer deposited under an angle of \SI{32}{\degree}. For the granular aluminium film we evaporate Al under an angle of \SI{0}{\degree} at a rate of \SI{2}{\angstrom\second^{-1}} while setting an oxygen pressure of $\SI{1.5e-5}{\milli\bar}$ inside the substrate chamber. The following two Al films for the Josephson junctions are evaporated under angles of \SI{\pm 17.5}{\degree} at a rate of \SI{3}{\angstrom\second^{-1}} with a pressure of $\SI{5e-8}{\milli\bar}$ inside the substrate chamber.

\subsection{Measurement setup}
The chip is placed in a copper holder box with an integrated superconducting coil to apply a magnetic field to change the magnetic flux threading the SQUID loop. This box is thermally anchored to the mixing chamber plate of a commercial dilution refrigerator with a base temperature of \SI{\sim 15}{\milli\kelvin}. To provide sufficient shielding the sample box is enclosed by copper and µ-metal shields, both anchored to the mixing chamber plate. 
The $IV$-curves are taken with compact DAQ modules from National Instruments, while the measurements of the $IR_\mathrm{diff}$-curves are measured with a Lock-in amplifier SR860 from Stanford Research Systems. The voltage drop across the SQUID is amplified by a factor of $10^4$ with a Femto DLPVA-100-F-D low-noise voltage amplifier.

\section{Data availability}

The data that support the findings of this study is available in Zenodo under \url{https://doi.org/10.5281/zenodo.11919737}.

\bibliography{Library}

\section{Ackowledgements}
We thankfully acknowledge useful discussions with A. B. Zorin, D. Scheer, F. Hassler, and M. Bieler. The authors also acknowledge technical support from H. Marx and N. Ubbelohde, and A. Fern\'andez Scarioni, M. Schröder, J. Blohm, P. Hinze and T. Weimann. This work was supported by the Deutsche Forschungsgemeinschaft (DFG) under Grant No. LO 870/2-1 and under Germany’s Excellence Strategy – EXC-2123 QuantumFrontiers – 390837967.

\newpage	
\newpage
\pagebreak
\newpage

\onecolumngrid

\newpage

\section{Supplementary}

\setcounter{section}{0}
\renewcommand{\thesection}{S\arabic{section}}
\setcounter{equation}{0}
\renewcommand{\theequation}{S\arabic{equation}}
\setcounter{table}{0}
\renewcommand{\thetable}{S\arabic{table}}
\setcounter{figure}{0}
\renewcommand{\thefigure}{\arabic{figure}}
\newcounter{SIfig}
\renewcommand{\theSIfig}{\arabic{SIfig}}

\newcounter{SIeq}
\renewcommand{\theSIeq}{S\arabic{SIeq}}

In this supplementary we provide further information on the experimental setup, the fabrication and characterization of the device, additional measurements complementing the measurements shown in the main text and further description on the numerical model.

\subsection{Experimental setup}

\begin{figure}[h]
	\includegraphics[width=0.75\textwidth]{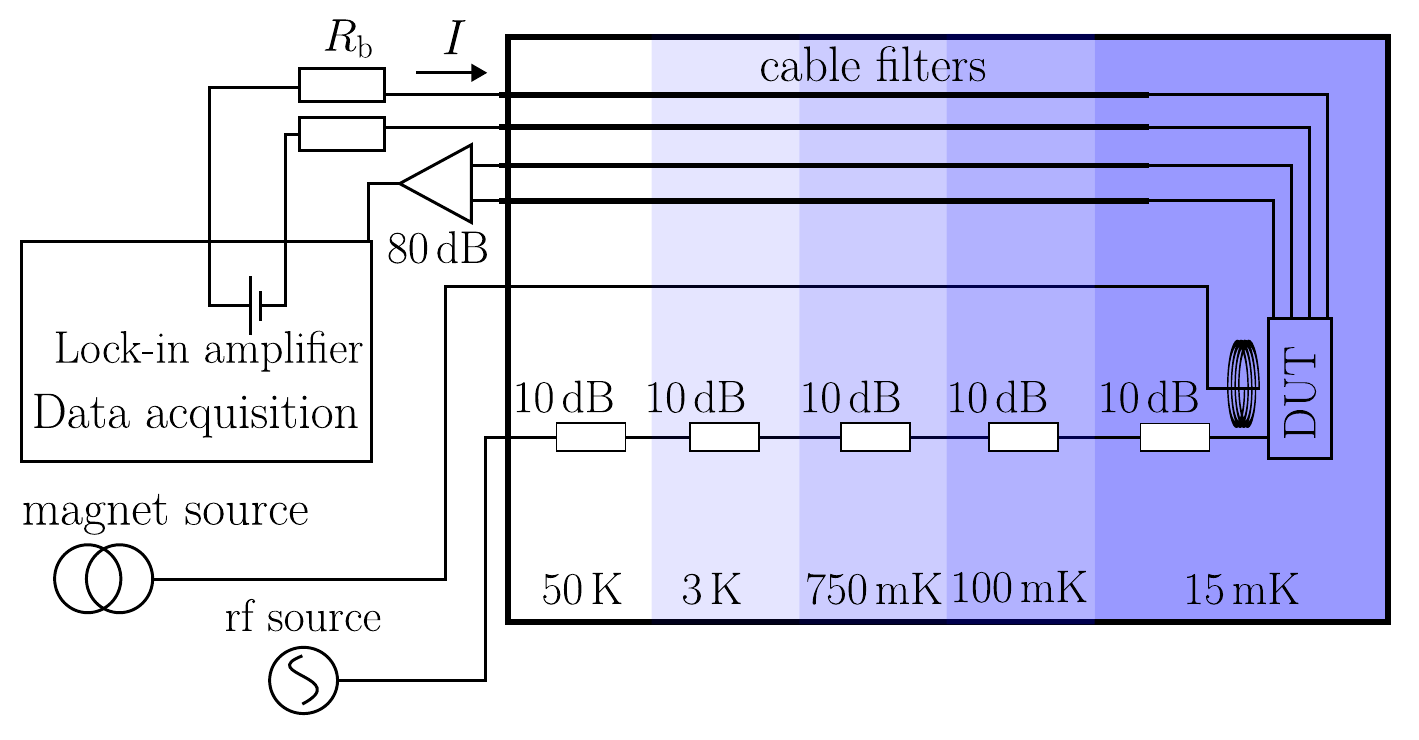}
	\caption{Schematics of the experimental setup. Different shades of blue indicate the temperature stages within the cryostat. See the text for detailed description of the used cabling and measuring schemes.}
	
	\label{cryostat}
\end{figure}

Supplementary Fig.~\ref{cryostat} depicts the experimental setup. The measurements were performed in an Oxford Instruments Triton 400 dilution refrigerator. The DUT is mounted in a rf-tight copper box, which has an integrated coil to apply the magnetic flux to the SQUID and thus, control $E_\mathrm{J,eff}$. The dc-lines from the sample box are connected to 3\:m long Thermocoax$\textsuperscript{TM}$ cable filters \citealp{Zorin95}, reaching from the mixing chamber stage ($\sim \SI{15}{\milli\kelvin}$) to the breakout connectors at the top of the cryostat. The $IV$-curves shown in Fig.~2 and Fig.~3(a) of the main text are measured by using CompactDAQ (cDAQ) converters from National Instruments. By applying dc-voltage $V_\mathrm{b}$ to two biasing resistors $R_\mathrm{b}=\SI{1}{\giga \ohm}$, a current $I=(V_\mathrm{b}-U)/(2R_\mathrm{b}+R_\mathrm{add})$ flows through the sample. $R_\mathrm{add}$ is the additional resistance from the cabling, the cable filters and the on-chip resistors. The voltage drop $U$ over the DUT is amplified by the low-noise voltage amplifier Femto DLPVA-100-F-D by a factor of $10^4$ (\SI{80}{\decibel}) and then digitalized using the cDAQ. For the measurements of differential resistance $R_\mathrm{diff}$ shown in Fig.~3(b-c) and Fig.~4(c-f) we use a Lock-in amplifier SR860 from Stanford Research Systems, where biasing resistors with \SI{100}{\mega\ohm}, the modulation frequency of \SI{13.97}{\hertz} and an amplitude of \SI{1}{\milli\volt} are used. The sinusoidal drives are provided by an Anapico APMS40G rf-generator and the arbitrary waveforms are generated by a Tektronix AWG7102. The rf-signals are attenuated by \SI{10}{\decibel} at each temperature stage of the cryostat, resulting in a total attenuation of $\SI{50}{\decibel}$, neglecting additional losses from cables and connectors. Measurements at room temperature yield an upper bound for the frequency dependent losses in the coaxial cables between room temperature electronics and rf input of the chip holder of $\lesssim \SI{1.7}{\decibel}$ at \SI{1}{\giga\hertz} and $\lesssim \SI{4.2}{\decibel}$ at \SI{6}{\giga\hertz}. To take the data presented in Fig.~4(e) the \SI{10}{\decibel} attenuator at the \SI{100}{\milli\kelvin} was removed, since the output voltage of the AWG was limited to \SI{1}{V}. The current for the flux coil is provided by either an Agilent B2901A SMU or by using an additional cDAQ channel.

\subsection{Equivalent circuit}

\begin{figure}[h]
	
	\includegraphics[width=0.65\textwidth]{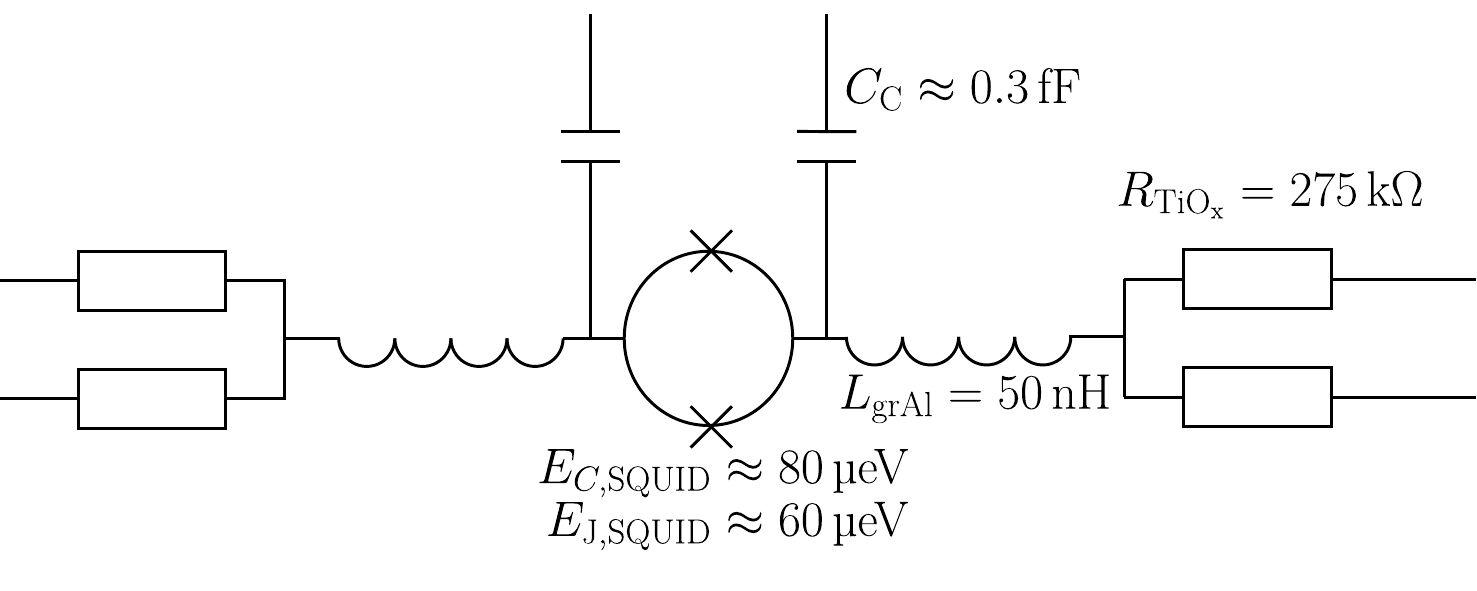}
	\caption{Equivalent circuit of the measured chip.}
	\label{Equivalent_circuit}
\end{figure}

To visualize the individual elements of the circuit Supplementary Fig.~\ref{Equivalent_circuit} shows the equivalent circuit of the sample shown in Fig.~1(c) of the main text.

\subsection{Fabrication techniques}

The granular aluminium and the titanium oxide layers were fabricated in the same electron beam evaporation system. For the grAl films, Al was evaporated with an evaporation rate of \SI{2}{\angstrom \second^{-1}}, while the wafer was placed in a chamber with an oxygen pressure of $p_\mathrm{O_2} \approx \SI{1.4e-5}{mbar} $. The same procedure was used for the TiO$_\mathrm{x}$, where titanium was evaporated at a rate of \SI{2}{\angstrom \second^{-1}} and the oxygen pressure was set to $p_\mathrm{O_2} \approx \SI{3.0e-6}{mbar} $. To ensure good electrical contact between the TiO$_\mathrm{x}$ and the following grAl and Al-layers, we sealed the end of the resistors with AuPd patches, which are less prone to oxidation in between the two fabrication steps.

\subsection{Characterization of the SQUID}

\begin{figure}[h]
	\renewcommand{\thefigure}{\arabic{figure}}	
	\centering\includegraphics[width=1\textwidth]{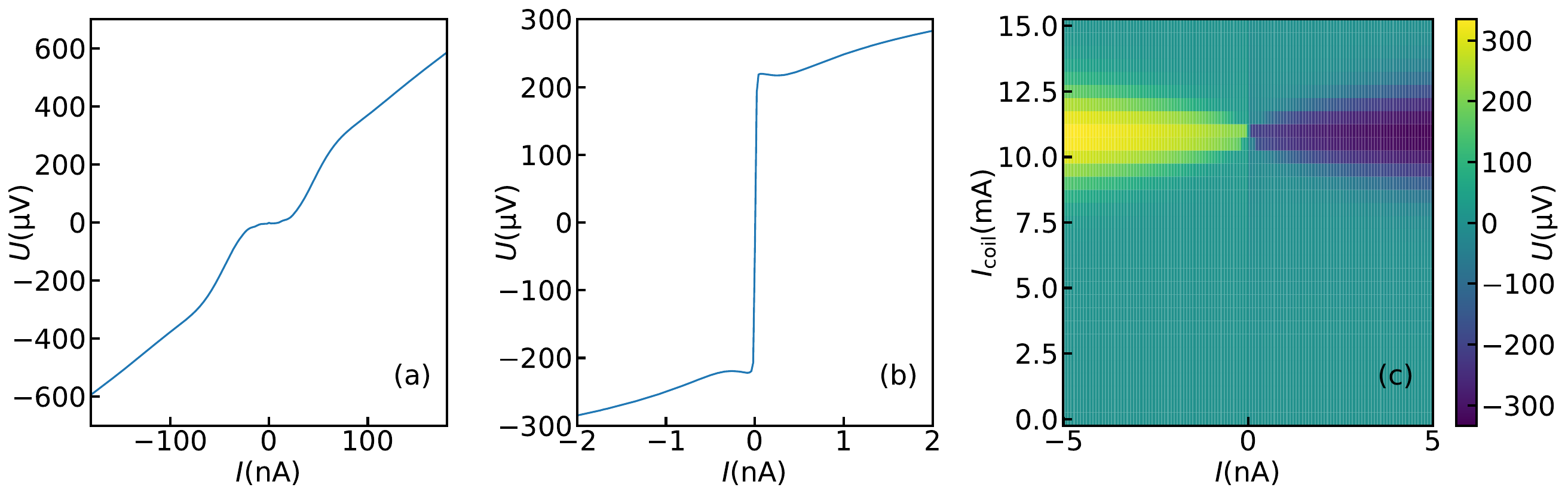}
	\caption{Characterization measurements: (a) Measured $IV$-curve for $\Phi=n \: \Phi_0$. From the critical current $I_c \approx \SI{30}{\nano \ampere}$ a Josephson energy $E_\mathrm{J} \approx \SI{60}{\micro \eV}$ can be estimated. (b) $IV$-curve at $\Phi=(n+1/2) \: \Phi_0$. We estimate the charging energy $E_C \approx \SI{80}{\micro \eV}$. (c) $IV$-curves for different coil currents $I_\mathrm{coil}$. $E_\mathrm{J}$ is maximally suppressed at $I_\mathrm{coil} \approx \SI{11}{\milli \ampere}$.}
	\label{characterization}
\end{figure}

Supplementary Fig.~\ref{characterization}(a) shows the $IV$-curves for $\Phi=n \: \Phi_0$, where the critical current $I_\mathrm{c} \approx \SI{30}{\nano\ampere}$, defined as the onset of the resistive branch, is maximized. The unsuppressed Josepshon energy is thus $E_\mathrm{J}=\frac{\Phi_0 I_\mathrm{c}}{2\pi} \approx \SI{60}{\micro\electronvolt}$. From the $IV$-curve in  Supplementary Fig.~\ref{characterization}(b) for $\Phi \to (n+1/2) \: \Phi_0$, where the Coulomb blockade is maximized, we can estimate the charging energy $E_C=\SI{80}{\micro\electronvolt}$. In Supplementary Fig.~\ref{characterization}(c) the $IV$-curves as a function of the coil current $I_\mathrm{coil}$ are shown. At $I_\mathrm{coil} \approx \SI{11}{\milli\ampere}$ the flux value $\Phi=\Phi_0/2$ is reached. From measurements up to $I_\mathrm{coil}=\SI{35}{\milli\ampere}$ we extract a current-to-flux coefficient of $\sim \SI{22.5}{\milli\ampere\Phi_0^{-1}}$.

\begin{figure}[h]
	\renewcommand{\thefigure}{\arabic{figure}}
	\centering\includegraphics[width=0.7\textwidth]{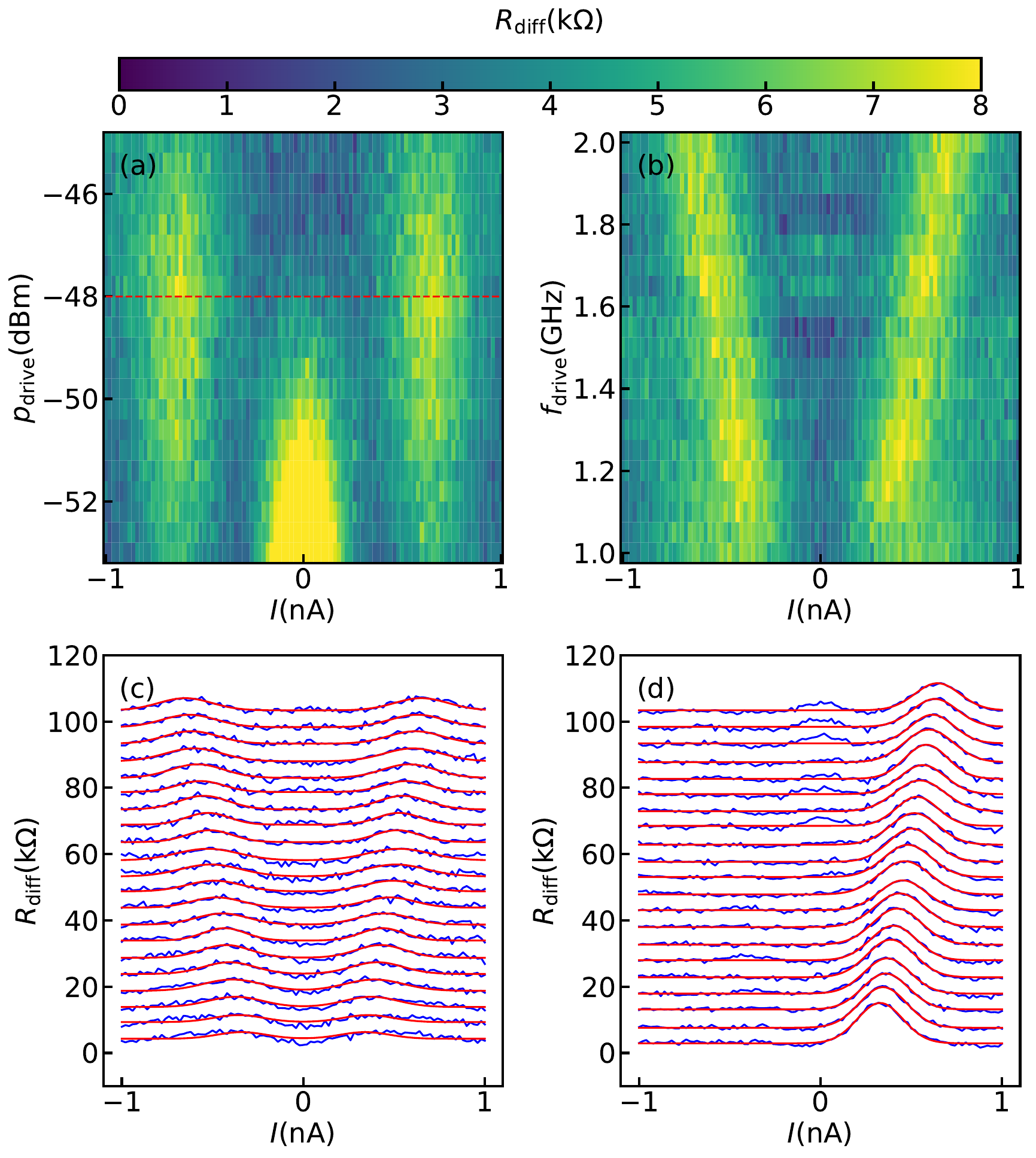}
	\caption{Detailed plots for Fig.~4 of the main text. (a) $R_\mathrm{diff}$ for different drive powers $p_\mathrm{drive}$. To find the best comparison with the measurement using a pulsed drive, we chose $p_\mathrm{drive}=\SI{-48}{\decibel m}$ (indicated with red line), at which the first dual Shapiro step is maximized. (b) $R_\mathrm{diff}$ for different frequencies $f_\mathrm{drive}$ of the sinusoidal drive. (c) Gaussian fits to the data shown in (b). For visualization purposes an offset of \SI{3}{\kilo \ohm} is added to each line. (d) Gaussian fits to the data shown in Fig.~4(f) of the main paper. An offset of \SI{5}{\kilo \ohm} is added to each line for clarity. The data shown in the inset of Fig.~4(f) are extracted from this measurement.}
	\label{Gauss}
\end{figure}

\subsection{Quantifying dual Shapiro steps}

The differential resistance $R_\mathrm{diff}$ for values between $\SI{-52.8}{\decibel m}\le p_\mathrm{drive}\le\SI{-44.2}{\decibel m}$ is shown in Supplementary Fig.~\ref{Gauss}(a). The red line at $\SI{-48}{\decibel m}$ indicates, where the measured peak $R_\mathrm{diff,peak}$ of the first dual Shapiro step is maximized and is used for comparison in Fig.~4(b) in the main text. To quantify the difference between the sinusoidal and the pulsed drive, we measured, analogue to Fig.~4(d) of the main text, the differential resistance for sinusoidal drive frequencies $\SI{1}{\giga \hertz}\le f_\mathrm{drive} \le \SI{2}{\giga \hertz}$ at $p_\mathrm{drive}=\SI{-48}{\decibel m}$, as shown in Supplementary Fig.~\ref{Gauss}(b). The peak values $R_\mathrm{diff,peak}$ shown in the inset of Fig.~4(d) of the paper are extracted by fitting Gaussians to the measurements of Fig.~\ref{Gauss}(b) and Fig.~4(d) of the paper. In Supplementary Fig.~\ref{Gauss}(c) and Supplementary Fig.~\ref{Gauss}(d) the fits for a set of frequencies are shown. Supporting the assumption that the smearing of the dual Shapiro steps occurs due to thermal broadening, fitting a Gaussian leads to good agreement with the measurements. The error bars in the inset of Fig.~4(f) of the main text are given by the error on the current of the dual Shapiro step.

\subsection{Numerical analysis of different drives}

\begin{figure}[h]
	\renewcommand{\thefigure}{\arabic{figure}}
	\centering\includegraphics[width=0.7\textwidth]{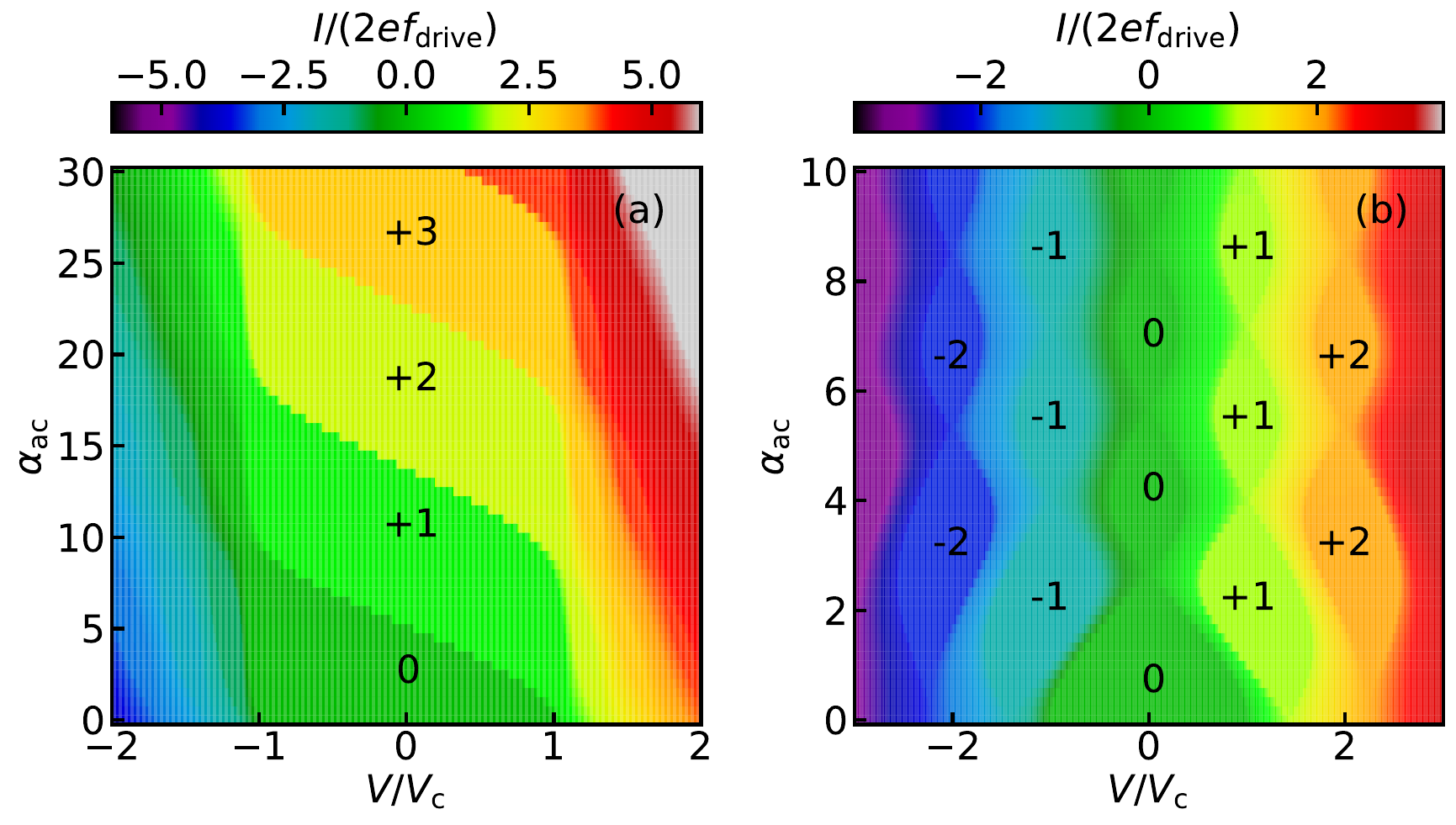}
	\caption{Simulated $IV$-curves for different drive powers obtained by solving Eq.~\eqref{diff} using the fourth-order Runge-Kutta method. (a) Pulsed drive with pulse duration $t_p= 0.05 \cdot f^{-1}_\mathrm{drive}$. The region 0 indicates the Coulomb blockade, while the numbers and sign indicate the order and sign of the dual Shapiro step. For a pulsed drive dual Shapiro steps only occur for one current direction. (b) Sinusoidal drive leads to a symmetric power dependence of the $IV$-curves and appearance of positive and negative dual Shapiro steps as indicated in the plot. }
	\label{simulation}
\end{figure}

To evaluate the influence of different drives on the dual Shapiro steps, we make use of the duality principle. For simplicity, we assume a sinusoidal shape of the lowest Bloch band and neglect all higher energy bands. With this single band approximation the dynamics of the quasicharge $q$ is described by
\begin{equation}
	\renewcommand{\theequation}{S.\arabic{equation}}
	L\ddot{q} + R \dot{q} + V_\mathrm{c}\sin\left( \frac{\pi}{e} q \right) = V_\mathrm{tot} = V_\mathrm{dc} + V_\mathrm{ac}(t).
	\label{diff_initial}	
\end{equation}
With $\tau = t\frac{\pi V_\mathrm{c}}{eR}$, $\beta= \frac{\pi V_\mathrm{c} L}{eR^2}$, $\tilde{q}=\frac{\pi}{e}q$ and $\alpha_\mathrm{dc,ac}= V_\mathrm{dc,ac}/V_\mathrm{c}$ the equation \eqref{diff_initial} simplifies to

\begin{equation}
	\renewcommand{\theequation}{S.\arabic{equation}}
	\beta \frac{d^2 \tilde{q}}{d\tau^2} + \frac{d\tilde{q}}{d \tau} + \sin \tilde{q} = \alpha_\mathrm{dc} + \alpha_\mathrm{ac}f(\tau)
	\label{diff}	
\end{equation}

and can be solved using a fourth-order Runge-Kutta method \cite{Maggi96}. Supplementary Fig.~\ref{simulation} shows the calculated $IV$-curves for different drive powers for a pulsed (a) and a sinusoidal (b) driving signal. For pulsed driving, the Coulomb blockade (colored region indicated by "0") gradually vanishes with increasing power and the first dual Shapiro step (colored region indicated by "+1") grows up to the point $\alpha_\mathrm{ac} \approx 10$, where the second dual Shapiro steps starts appearing. The negative dual Shapiro steps never appear in this calculation due to the sign of the current pulse. Changing the sign of the pulse would lead to a similar pattern of steps at negative currents. The measured power dependence shown in Fig.~4(d) of the main paper qualitatively supports this calculation. For a sinusoidal drive shown in Supplementary Fig.~\ref{simulation}(b), increasing the power leads to a suppression of the Coulomb blockade and an appearance of both positive and negative steps symmetrically until at $\alpha_\mathrm{ac}=2.5$ the CB completely vanishes. Further increasing the power leads to a reappearing of the Coulomb blockade and a pronounced second dual Shapiro step. This pattern is matching the measurement shown in Fig.~3(c) in the main text.

\subsection{Temperature dependence of the Coulomb blockade}

\begin{figure}[h]
	\renewcommand{\thefigure}{\arabic{figure}}
	\centering\includegraphics[width=0.8\textwidth]{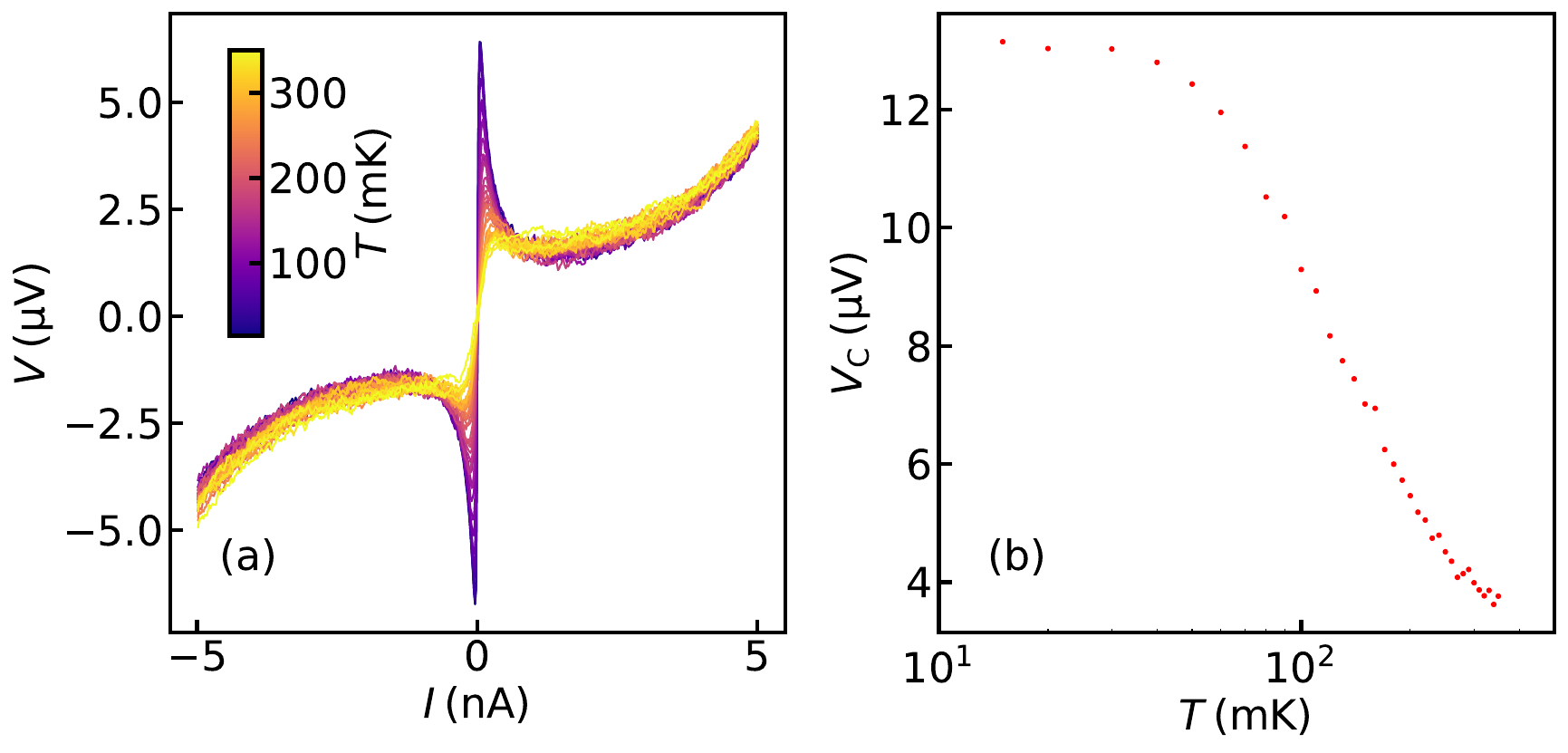}
	\caption{(a) $IV$-curves of the sample at $\Phi \approx 0.3\Phi_0$ at different temperatures (indicated by the color). (b) Extracted Coulomb blockade $V_\mathrm{c}$ as a function of the temperature. }
	\label{Coulomb_temperature}
\end{figure}

Supplementary Figure~\ref{Coulomb_temperature}(a) shows the temperature dependence of the $IV$-curves at an external flux of $\Phi \approx 0.3\Phi_0$. One can see that the Coulomb blockade is shrinking with increasing temperature. The extracted voltage of the Coulomb blockade $V_\mathrm{c}$ is shown in Supplementary Fig.~\ref{Coulomb_temperature}(b). The Coulomb blockade shows a small plateau up to \SI{40}{\milli\kelvin} and then starts to shrink with increasing temperature. This indicates that without rf-drive the electron temperature is close to the fridge temperature. To get an estimate of the electron temperature during the irradiation of the rf-drive, one can compare the $IV$-curve of Fig.~3(a) to the one of Supplementary Fig.~\ref{Coulomb_temperature}. We thus expect the electron temperature during the rf-drive to be around \SI{200}{\milli\kelvin} to \SI{250}{\milli\kelvin}.

\subsection{Choice of flux point}

To investigate the effect of the flux bias point on the dual Shapiro steps the current of the coil generating the magnetic field was varied, while the rf drive was fixed at a power $p_\mathrm{drive}=\SI{-48}{dBm}$ and frequency $f_\mathrm{drive}=\SI{3.2}{\giga\hertz}$ and the $IV$-curves were measured. To find a good trade-off between the width of the dual Shapiro step and the Landau-Zener tunneling branch, the  differential resistance normalized to the average of the normal resistance $\bar{R}_\mathrm{diff}$ was plotted in Supplementary Fig.~\ref{Vary_coil_dSs} (a). It can be seen, that a small coil current $I_\mathrm{coil}$, the dual Shapiro steps can not be seen due to the noise of the measurement (red). Increasing $I_\mathrm{coil}$ makes the dual Shapiro step visible (orange). Further increasing $I_\mathrm{coil}$ increases the width of dual Shapiro step but at the same time the Landau-Zener tunneling, such that the normalized $R_\mathrm{diff}$ shows a less pronounces dual Shapiro step (magenta). The $IV$-curves of the three coil currents indictated by the colored lines are shown in Supplementary Fig.~\ref{Vary_coil_dSs}(b). The orange curve shows the clearest dual Shapiro step, while the red curve is dominated by the noise of the measurement and the magenta curve is showing a less pronounced step when comparing it to the voltage increase due to Landau-Zener tunneling.

\begin{figure}[h]
	\renewcommand{\thefigure}{\arabic{figure}}
	\centering\includegraphics[width=0.8\textwidth]{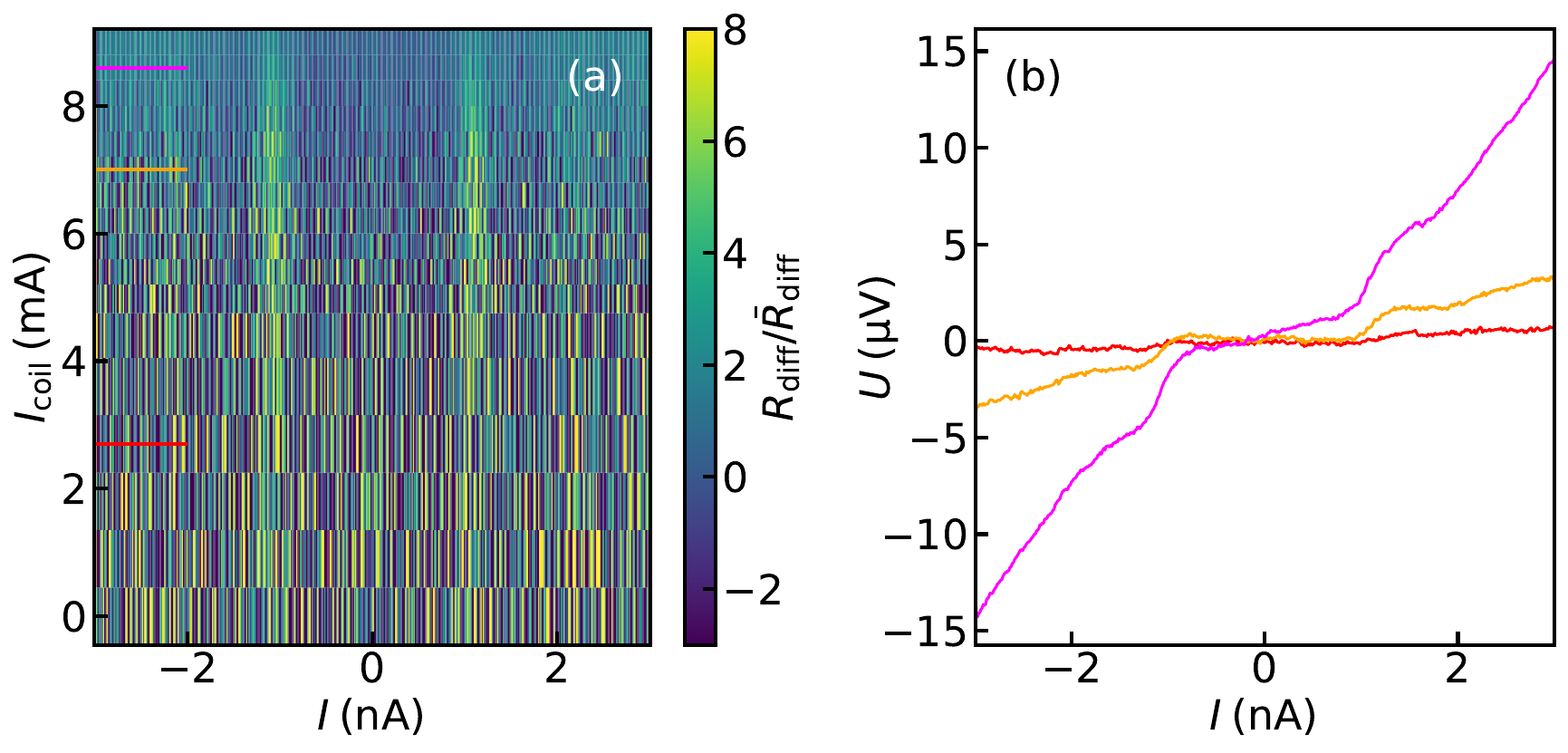}
	\caption{(a) Differential resistance extracted from measuring the $IV$-curves and normalized to the average differential resistance to take the Landau-Zener tunneling into account. (b) $IV$-curves of the three colored lines in (a). }
	\label{Vary_coil_dSs}
\end{figure}

\end{document}